\documentclass{JHEP3}
\pdfoutput=1
\usepackage{epsfig}

\newcommand{\be}{\begin{equation}}
\newcommand{\ee}{\end{equation}}
\newcommand{\bea}{\begin{eqnarray}}
\newcommand{\eea}{\end{eqnarray}}

\newcommand{\tr}{{\rm tr\ }}

\title{
Introduction of the chemical potential in the overlap formalism
}
\author{R. Narayanan\\
Department of Physics, Florida International University, Miami,
FL 33199.
\\E-mail: \email{rajamani.narayanan@fiu.edu}}
\author{Sayantan Sharma\\
Department of Theoretical Physics, Tata Institute of Fundamental
Research,
Homi Bhabha Road, Mumbai 400005, India
\\E-mail: \email{ssharma@theory.tifr.res.in}
}

\abstract{
We investigate the possibility of coupling a chemical potential only
to the physical chiral fermions on the lattice starting from 
the many body state description of overlap fermions.
After developing the formalism for a chiral gauge theory, we focus
our attention on the case of free fermions coupled to a vector like
chemical potential and discuss the issue of zero temperature divergences.
}


\preprint{TIFR/TH/11-36}

\begin{document}

\section{Introduction}

Nonperturbative studies of QCD at finite baryon density and finite
temperature is an active field of investigation on the lattice with
implications for current experimental studies of the transition
between the hadronic phase and quark-gluon plasma 
phase~\cite{Endrodi:2011gv,Gupta:2009mu,Gavai:2008zr,Ejiri:2005wq}.
Due to the well known ``sign-problem'' associated with lattice
simulations
at finite chemical potential, a large fraction of the numerical
simulations
compute coefficients 
in the Taylor expansion
in chemical potential~\cite{Allton:2002zi,Gavai:2003mf}. Typically one computes
susceptibilities (second order coefficients) but higher moments have
also been computed~\cite{Gavai:2008zr,DeTar:2010xm}. The Taylor expansion is
expected
to diverge close to the critical point where fluctuations dominate the
physical
behavior~\cite{Asakawa:2000wh}. 

We will focus on 
the quark number susceptibility for definiteness in this paper but
the formalism in this paper carries over to all coefficients in a
Taylor expansion.
Let $Z(\mu,N_T)$ be the grand
canonical partition
function
on a $N^3\times N_T$ lattice at a given chemical potential $\mu$.
The dependence of the partition function on other quantities like
gauge coupling, quark masses etc. have been suppressed. We also
assume that we will take the thermodynamic limit, $N\to\infty$,
and $N_T^{-1}$ will play the role of temperature on the lattice.
The quark number susceptibility is defined as
\be
\chi(N_T) = \lim_{N\to\infty} \frac{1}{N^3 N_T} \frac{\partial^2 \ln
  Z(\mu,N_T)}{\partial \mu^2}\Bigg|_{\mu=0}.
\ee
This quantity has been studied both on the
lattice~\cite{Endrodi:2011gv,Gavai:2005rf}
and in other non-perturbative approaches~\cite{Blaizot:2002xz}.

A n\"aive
introduction of the number density operator on the lattice leads
to divergences at zero temperature in the following sense.
A proper definition should yield a result that goes as
$\frac{1}{N_T^2}$ and the presence of a non-zero $\chi(\infty)$
implies a divergence in the continuum limit.
Such a divergence
can be removed by replacing
$U_4(x)$ and $U^\dagger_4(x)$ on the lattice by $e^\mu U_4(x)$
and $e^{-\mu} U_4^\dagger(x)$ 
where $U_4(x)$ is the link variable on the lattice
connecting $x$ and $x+\hat 4$~\cite{Hasenfratz:1983ba} and will be
referred as the Hasenfratz-Karsch (H-K) prescription. Let
$T_4$ denote operator
\be
T_4 \psi(x) = U_4(x) \psi(x+\hat 4).\label{t4op}
\ee
For n\"aive fermions, the source term added to the action would be
\be
j_4 = \cases { 
\bar\psi \gamma_4 \psi & for
   n\"aive insertion;\cr
\bar\psi \gamma_4\frac{(e^\mu-1)T_4-(e^{-\mu}-1)T_4^\dagger}{2}\psi
& for
   H-K insertion. \cr }
\ee
The n\"aive insertion would result in a non-zero $\chi(\infty)$ but
the in H-K insertion it would lead off as $\frac{1}{N_T^2}$.

The aim of this paper is to address the introduction of the
chemical potential into the overlap Dirac
operator~\cite{Neuberger:1997fp}. 
It was introduced
using the H-K prescription by Bloch and Wettig~\cite{Bloch:2006cd}.
The Wilson-Dirac operator, $H_w(\mu)$, that appears as the kernel in
the definition of the overlap Dirac operator is not Hermitian in the
presence
of a chemical potential. The massless overlap Dirac operator in the
presence
of the chemical potential as per the H-K prescription is
\be
D_{\rm BW}(\mu) = \frac{1+ \gamma_5 \epsilon(H_w(\mu))}{2}.\label{dbw}
\ee
Assuming $H_w(\mu)$ is diagonalizable in the form
\be
H_w(\mu) = R \Lambda R^{-1},
\ee
with $\Lambda$ being a diagonal matrix with complex entries, one sets
\be
\epsilon(H_w(\mu)) = R \epsilon ({\rm Re}\Lambda) R^{-1}.
\ee
Such a choice can be justified using the domain wall
formalism~\cite{Bloch:2007xi}
since $\epsilon(H_w)$ is the limit of $\tanh(L_s H_w)$ with
$L_s$ being the extent in the extra continuous dimension~\cite{Neuberger:1997bg}.
In spite of the fact that $V(\mu)=\gamma_5 \epsilon(H_w(\mu))$ is not a
unitary
operator for $\mu\ne 0$, the massless propagator defined in the 
usual manner~\cite{Neuberger:1997bg,Edwards:1998wx} as
\be
G(\mu) = \frac{1-V(\mu)}{1+V(\mu)}
\ee
is still chiral: Let us write
\be
R = \pmatrix{\alpha & \gamma \cr \beta & \delta \cr};\ \ \ 
\Lambda = \pmatrix{\Lambda^+ & 0 \cr 0 & -\Lambda^-\cr};\ \ \ {\rm Re}
  \Lambda^{\pm} > 0.
\ee
Then one can show that
\be
G(\mu) = \pmatrix{0 & \gamma\delta^{-1} \cr \beta\alpha^{-1} & 0 \cr},
\ee
showing the chiral nature of the massless propagator assuming that
$\alpha$ and $\delta$ are invertible complex matrices.
Both the energy density~\cite{Gattringer:2007uu} 
and
the quark number susceptibility~\cite{Banerjee:2008ii} for the case of free overlap fermions
were analyzed within the Bloch-Wettig scheme and shown to have the
correct
continuum behavior.
One possible objection to the Bloch-Wettig scheme would be that the
chemical
potential is coupled to all the fermions from the view-point of domain
wall
fermions and not just to the chiral fermions. This objection could
also
be raised for Wilson fermions where the chemical potential is coupled
to the massless mode as well as the massive doublers but we should
keep in mind that
we have infinite number of fermions in the case of domain wall fermions.
Since we have an explicit operator for the physical fermions, it
should be possible to
couple
the chemical potential just to the physical fermions in the overlap formalism.
Another possible objection already raised in~\cite{Bloch:2006cd} is
that
the number of topological modes can depend on the chemical potential
since the topological charge is 
\be
Q = \int d^4 x q(x) = \frac{1}{2}\int d^4 x \tr \epsilon (x),
\ee
where $\tr$ is the trace over the spin components at a fixed point
$x$.
Since $\epsilon(H_w(\mu))$ depends on $\mu$, $Q$
can change with $\mu$ in the Bloch-Wettig scheme. 
In fact $q(x)$ was shown to depend on $\mu$ in~\cite{Gavai:2009vb}.

We will start from the
first principles of the overlap formalism~\cite{Narayanan:1994gw} and add a chemical
potential by coupling it to the number density operator. 
We will consider the general case where the chemical potential
for the fermions with positive and negative chirality can be
different.
We will
make a class of choices for the number density operator all of
which reduce to the standard continuum limit. We will focus on the
case of free overlap Dirac fermions and show that divergences in the
quark
number susceptibility is a generic feature. Since every member in our
class of choices will contain the correct $\frac{1}{3N_T^2}$
in the free quark number susceptibility, we will be able to write down
a large class of number density operators that will not contain a
divergent
piece. 

As a first step, we will review the generating functional~\footnote{We
  will only consider the generating functional for a fixed gauge field
  background
through out this paper.}
for a
massless
vector like gauge theory. None of this is new and can be found in one
form or other in~\cite{Neuberger:1997bg,Narayanan:1994gw}. In particular,
we will explicitly write down the chiral transformations in the overlap 
formalism and derive the massless overlap
Dirac
operator~\cite{Neuberger:1997fp}. This will help us
introduce the chemical potential and derive the associated
generating functional for the massless overlap Dirac operator with
a chemical potential. The resulting class of operators will be different
from $D_{\rm BW}(\mu)$ in (\ref{dbw}) but the generating functional will be invariant
under chiral transformations. All these operators will have two
features that separate them from the operator in the Bloch-Wettig
scheme: Chemical potential will only be coupled to the physical chiral
fermions and topological charge will not depend on the chemical
potential.
Finally, we will discuss the addition of a mass term to the generating functional
that destroys the chiral symmetry.
We will end the paper with a discussion of the quark number
susceptibility for free overlap Dirac fermions arising from our
generating functional.

\section{The overlap formalism}

This section is essentially a repeat of what can be found
in~\cite{Narayanan:1994gw}. The main difference is that we only use
one Hermitian Wilson-Dirac operator in the construction of overlap
fermions.
This amounts to setting the Wilson mass parameter to infinity on the
inactive side. This changes the details as it pertains to the
generating functional and since we will derive the generating
functional
for massive overlap fermions in the presence of the chemical
potential,
we will need explicit results from this section.

At the heart of the overlap formalism is the massive Hermitian Dirac
operator which is usually realized on the lattice using the
Wilson-Dirac operator, $H_w$. The notational details are given in Appendix~\ref{ap1}.
 The generating functional for a vector like theory with massless
fermions
is given by~\cite{Narayanan:1994gw}
\be 
Z(\bar\xi,\xi)
={}_R\langle-|e^{\bar\xi_R d_R+\xi_R u^\dagger_R}|+\rangle_R \ \ 
{}_L\langle +|e^{\xi_L d^\dagger_L +\bar\xi_L u_L}|-\rangle_L.
\label{genfun1}
\ee 
Dirac valued operators, $a_{R,L},a_{R,L}^\dagger$, obey canonical 
anti-commutation relations separately for the $R$ and $L$ sets. 
We write,
\be
a_{R,L}=\pmatrix{u_{R,L}\cr d_{R,L}\cr},\label{baweyl}
\ee
in terms of their respective Weyl components.
The sources, 
\be
\bar\xi = \pmatrix{\bar\xi_R &\bar\xi_L\cr};\ \ \ \ 
\xi = \pmatrix{\xi_R\cr\xi_L\cr}
\ee
are Dirac values
Grassmann variables  that
couple
directly
to the physically relevant fields
and they anti-commute with the fermionic operators.
$|+\rangle_{R,L}$ are lowest states of
\be
{\cal H}_{R,L} = -a_{R,L}^\dagger H_w a_{R,L}
\ee 
and $|-\rangle_{R,L}$ are lowest states of
two many body operators,
\footnote{The pair ${\cal H}_R$ and
 $\Gamma_R$
replaces the ${\bf H}_\pm$ in~\cite{Narayanan:1994gw}. That one can
take the Wilson mass parameter to infinity on one side follows from
the discussion of the topological charge in section 8 of~\cite{Narayanan:1994gw}.}
\be
{\Gamma}_{R,L} = -a_{R,L}^\dagger \gamma_5 a_{R,L}.
\ee

We now make several remarks on the generating functional.
\begin{enumerate}
\item The phase choice for $|+\rangle_{R,L}$ are tied together since
  they are the ground states of identical many body operators.
The same is true for $|-\rangle_{R,L}$. Therefore, the generating functional
is unambiguous and does not not depend upon the
phase choice present in the unitary matrix, $U$, that diagonalizes $H_w$.
\item
It does not depend
on the ordering of the operators 
since the two terms in the exponent commute with each
other
in both factors.
\item
It is clear that $d_R$ and
$u^\dagger_R$ are the propagating degrees of freedom in the first factor
since ${}_R\langle-| u_R$ and ${}_R\langle-| d^\dagger_R$ are both zero.
The converse holds for the second factor. 
\item
The generating functional is invariant under global chiral
transformations:
\be
\xi_R \to e^{i\varphi_R} \xi_R;\ \ 
\bar\xi_R \to \bar\xi_R e^{-i\varphi_R} ;\ \ 
\xi_L \to e^{i\varphi_L} \xi_L;\ \ 
\bar\xi_L \to \bar\xi_L e^{-i\varphi_L} .\label{chiral}
\ee
\item 
The explicit expression for the generating functional can
be obtained by small modifications to the details presented
in~\cite{Narayanan:1994gw}
and the result is
\be
Z(\bar\xi,\xi)=
\left[e^{\bar\xi_R \beta\alpha^{-1} \xi_R}
\det\alpha \right]
\left[e^{\bar\xi_L \left[\beta\alpha^{-1}\right]^\dagger \xi_L}
\det\alpha^\dagger\right].
\label{genfun}
\ee
In particular,
the propagators for right-handed and left-handed fermions are
\be
G^{ij}_R = \frac{{}_R\langle - | {u^\dagger_R}_j {d_R}_i
 |+\rangle_R}{{}_R\langle-|+\rangle_R} =\left[\beta\alpha^{-1}\right]_{ij};\ \ \ 
G^{ij}_L = \frac{{}_L\langle + | {d^\dagger_L}_j {u_L}_i
 |-\rangle_L}{{}_L\langle +|-\rangle_L}=\left[\beta\alpha^{-1}\right]^\dagger_{ij},
\ee 
and they obey the relation
\be
G^\dagger_R = G_L.
\ee
\end{enumerate}

In practice, one can avoid exact diagonalization of $H_w$ which is
needed for the computation of $U$ and (\ref{genfun}) since one has
an overlap-Dirac operator for vector like
theories~\cite{Neuberger:1997fp}.
Consider the unitary operator,
\be
V = \gamma_5 \epsilon(H_w).
\ee
It follows from (\ref{Hweig}) and (\ref{umat}) that
\be
\frac{1+V}{2} U = \pmatrix { \alpha & 0  \cr 0 & \delta \cr};\ \ \ 
\frac{1-V}{2} U = \pmatrix { 0 & \gamma \cr \beta & 0 \cr},\label{pmv}
\ee
and therefore
\be
G_o=\frac{1-V}{1+V} = \pmatrix{ 0 &-\left(\beta\alpha^{-1} \right)^\dagger\cr 
\beta\alpha^{-1} & 0  \cr},\label{ferprop}
\ee
is the massless overlap Dirac propagator.
Since~\cite{Narayanan:1994gw}
\be
\det U = \frac{\det\alpha}{\det\delta^\dagger};
\ee
we have the identity,
\be
\det \alpha \det\alpha^\dagger = \det\delta \det \delta^\dagger,
\ee
and therefore,
\be
\det \frac{1+V}{2} = \det\alpha\det\alpha^\dagger.\label{ferdet}
\ee
The massless overlap Dirac operator is given by
\be
D_o = \frac{1+V}{2}.\label{ovop}
\ee
Our generating functional in (\ref{genfun}) can be written
as~\footnote{The presence of $-i\gamma_4$ is due to our choice of
  basis for the chiral sources.}
\be
Z(\bar\xi,\xi)=\det D_o e^{-i\bar\xi\gamma_4 G_0\xi }.
\ee
We can use (\ref{ferprop}) and (\ref{ferdet}) along with an efficient
implementation
of $V$ to compute the generating functional.
Note that the operator in (\ref{ovop})
is not identical to the operator used to compute the propagator in
(\ref{ferprop}).
This is an essential ingredient of the overlap formalism. 

\section{Introduction of the chemical potential}\label{chemsec}

Consider the generating functional
\be 
Z(\bar\xi,\xi;\hat\mu_R,\hat\mu_L) =
{}_R\langle-|
e^{\bar\xi_R d_R+ \xi_R u^\dagger_R+ u^\dagger_R
\hat\mu_R d_R}|+\rangle_R 
{}_L\langle +|e^{\xi_L d^\dagger_L  +\bar\xi_L u_L - d^\dagger_L
\hat\mu_L u_L}
|-\rangle_L,\label{genchem}
\ee 
where $\hat\mu_R(\mu_R)$ and $\hat\mu_L(\mu_L)$ are operators that
parametrically depend on the chemical potentials, $\mu_R$ and $\mu_L$.

In order to obtain a formula for the generating functional,
we start by noting that
\be
\int   
d\zeta_R d\bar\zeta_R e^{-\bar\zeta_R \hat\mu_R d_R - 
\zeta_R u^\dagger_R + \bar\zeta_R\zeta_R} = e^{u^\dagger_R
\hat\mu_R d_R}
\ee  
and
\be
\int  
d\bar\zeta_L d\zeta_L e^{-\bar\zeta_L \hat\mu_L u_L - 
\zeta_L d^\dagger_L - \bar\zeta_L\zeta_L} = e^{- d^\dagger_L
\hat\mu_L u_L},
\ee 
where $\zeta_R,\zeta_L,\bar\zeta_R,\bar\zeta_L$ are Grassmann
variables that anticommute with all fermionic operators and Grassmann variables.
Therefore (\ref{genfun1}) and (\ref{genfun}) gets modified to
\bea
Z(\bar\xi,\xi;\mu_R,\mu_L)
&=& \det \alpha \det \alpha^\dagger\cr
&&
\int d\zeta_R d\bar\zeta_R e^{ \bar\zeta_R\zeta_R + 
\left(\bar\xi_R-\bar\zeta_R\hat\mu_R\right)
\beta\alpha^{-1} 
\left(\xi_R-\zeta_R \right)} \cr
&&
\int  d\bar\zeta_L d\zeta_L e^{- \bar\zeta_L\zeta_L  + 
\left(\bar\xi_L -\bar\zeta_L\hat\mu_L \right)
\left[\beta\alpha^{-1}\right]^\dagger 
\left(\xi_L-\zeta_L\right)}\cr
&=& 
\det \alpha \det \left( 1+ \hat\mu_R \beta\alpha^{-1}\right)
\det \alpha^\dagger \det \left( 1 -\hat\mu_L
 \left[\beta\alpha^{-1}\right]^\dagger \right)\cr
&& 
e^{\bar\xi_R \frac{1}{\alpha\beta^{-1}+\hat\mu_R}\xi_R}
 e^{\bar\xi_L
 \frac{1}{\left[\alpha\beta^{-1}\right]^\dagger-\hat\mu_L}\xi_L}~.
\label{genchem1}
\eea

Note that the generating functional is invariant under the chiral
transformation
given in (\ref{chiral}) since the introduction of the chemical
potential does not mix the two chiral sectors. To be consistent
with the continuum definition of the chemical potentials,
we require 
\be
\hat\mu_R(0) = 0;\ \ \ \ 
\hat\mu_L(\mu) = -\hat\mu^\dagger_R(-\mu).
\ee

Note that the
fermion determinant is not real and positive for a real quark
chemical potential, $\mu_R=\mu_L=\mu$,
but it is real and positive for an isospin chemical potential,
$\mu_R=-\mu_L=\mu$.
These are standard properties in the continuum that are correctly
reproduced by
the overlap formalism.

There are several options for $\hat\mu_R(\mu)$. 
We will address this issue when we analyze free fermions.

We can write (\ref{genchem1}) in a compact form by introducing the
massless overlap Dirac operator and propagator in the
presence
of a chemical potential.
If we define
\bea
N &=& \pmatrix{ 0 & \hat\mu_R \cr \hat\mu_L & 0 \cr};\cr
D_o(\hat\mu_R,\hat\mu_L) &=& \frac{1+V}{2} + N \frac{1-V}{2} \cr
G_o(\hat\mu_R,\hat\mu_L) &=& \left[ \frac{1+V}{1-V} + N \right]^{-1} .\label{ovop1}
\eea
then a straight forward computation shows that
\be
Z(\bar\xi,\xi;\hat\mu_R,\hat\mu_L) =
\det D_o(\hat\mu_R,\hat\mu_L) e^{-i\bar\xi \gamma_4 G_o(\hat\mu_R,\hat\mu_L)\xi}.
\ee

\section{Introduction of the fermion mass}

A mass term can be added in analogy with the chemical potential
term~\cite{Narayanan:1994gw}.
The generating functional
in the presence of a chemical 
potential and mass term is defined to be
\bea
Z(\bar\xi,\xi;\hat\mu_R,\hat\mu_L;m) &=&
\left[ {}_R\langle-|\otimes {}_L\langle +| \right]\cr
&&e^{\bar\xi_R d_R+ \xi_R u^\dagger_R+ u^\dagger_R
 \hat\mu_R d_R+
\xi_L d^\dagger_L  +\bar\xi_L u_L - d^\dagger_L
 \hat\mu_L u_L +md^\dagger_L d_R - m u^\dagger_R u_L
}\cr
&&\left[ |-\rangle_L \otimes |+\rangle_R \right]\label{genfunmc1}
\eea
One can show using methods similar to one in (\ref{chemsec}) (see
Appendix~\ref{ap2} for details)
that
\bea
&&Z(\bar\xi,\xi;\hat\mu_R,\hat\mu_L;m)\cr
&=&
\det \alpha \det \left( 1+ \hat\mu_R \beta\alpha^{-1}\right)
\det \alpha^\dagger \det \left( 1 -\hat\mu_L
\left(\beta\alpha^{-1}\right)^{\dagger}\right) \cr
&&\det\left[ 1 + m^2 \frac{1}{\alpha\beta^{-1}+\hat\mu_R}
\left[
\frac{1}{\left(\alpha\beta^{-1}\right)^\dagger-\hat\mu_L}\right]\right]\cr
&&
e^{\bar\xi_R 
\left(\left(\alpha\beta^{-1}\right)^\dagger-\hat\mu_L\right)
\frac{1}{m^2+ \left(\alpha\beta^{-1}+\hat\mu_R\right)
\left(\left(\alpha\beta^{-1}\right)^\dagger-\hat\mu_L\right)}
\xi_R+
\bar\xi_L 
\left(\alpha\beta^{-1}+\hat\mu_R\right)
\frac{1}{m^2+ \left(\left(\alpha\beta^{-1}\right)^\dagger-\hat\mu_L\right)
\left(\alpha\beta^{-1}+\hat\mu_R\right)}
\xi_L}\cr
&&e^{\bar\xi_R
\frac{m}{m^2+ \left(\left(\alpha\beta^{-1}\right)^\dagger-\hat\mu_L\right)
\left(\alpha\beta^{-1}+\hat\mu_R\right)}
\xi_L
-\bar\xi_L
\frac{m}{m^2+ \left(\alpha\beta^{-1}+\hat\mu_R\right)
\left(\left(\alpha\beta^{-1}\right)^\dagger-\hat\mu_L\right)}
\xi_R}.
\label{genfunmc}
\eea

We can write (\ref{genfunmc}) in a compact form by extending the
massless overlap Dirac operator in the presence of a chemical
potential define in (\ref{ovop1}) to include a mass as follows:
If we define
\bea
N(m) &=& \pmatrix{ m & \hat\mu_R \cr \hat\mu_L & m \cr}\cr
D_o(\hat\mu_R,\hat\mu_L;m) &=& \frac{1+V}{2} + N(m) \frac{1-V}{2} \cr
G_o(\hat\mu_R,\hat\mu_L;m) &=& \left[ \frac{1+V}{1-V} + N(m) \right]^{-1} .\label{ovopm}
\eea
then a straight forward computation show that
\be 
Z(\bar\xi,\xi;\hat\mu_R,\hat\mu_L;m) =
\det D_o(\hat\mu_R,\hat\mu_L;m) e^{-i\bar\xi \gamma_4 G_o(\hat\mu_R,\hat\mu_L;m)\xi}.
\ee 
Note that the expressions reduce to the usual ones~\cite{Edwards:1998wx} for the massive
overlap Dirac operator without a chemical potential.

\section{Free overlap fermions}

In order to work out the energy density and quark number
susceptibility for free massless quarks, it is best to work in
momentum space. Let us assume that we have converted to creation and
annihilation operators in momentum space by the appropriate unitary
transformation. We will work on a $\infty^3 \times N_T$ lattice.
The allowed spatial momenta are $p_k\in [-\pi,\pi]$.
The Matsubara frequencies in the $N_T$ direction are $\omega_n=\pm\frac{2\pi n +
 \pi}{N_T}$; $n=0,\cdots, \frac{N_T}{2}-1$ and we will assume that $N_T$ is even.
The hermitian Wilson-Dirac operator for a fixed momentum takes the
form
\be
H_w = \pmatrix{w & c \cr c^\dagger & -w \cr}
\ee
where
\bea
w=2\sum_k \sin^2\frac{p_k}{2} +2\sin^2\frac{\omega_n}{2}- m_w;&& 
c=i\sum_k \sigma_k \sin p_k - \sin\omega_n;\cr cc^\dagger = c^\dagger c &=&
\sum_k \sin^2 p_k + \sin^2\omega_n.
\eea
The positive eigenvalues come in one doubly degenerate 
pair, $\lambda=\sqrt{w^2+cc^\dagger}$, per momentum block and the corresponding pair of 
orthonormal eigenvectors are
\be
\frac{1}{\sqrt{2\lambda (\lambda-w)}}\pmatrix{c \cr \lambda-w
 \cr} .
\ee 
Note that $\lambda$ and $w$ are even functions of $\omega_n$ at a
fixed $p_k$.
Assuming a vector like chemical potential, 
we write
\be
N = \pmatrix{ 0 & n(\mu,\omega_n) \cr
-n^*(-\mu,\omega_n) & 0 \cr},
\ee
per momentum block where $n(0,\omega_n)=0$.
We write,
\be
n(\mu,\omega_n) = n_r(\mu,\omega_n)+in_i(\mu,\omega_n)
\ee
where $n_{r,i}(\mu,\omega_n)$ are both real functions.
The free energy density is given by 
\be
\ln Z(\mu,N_T) = 
\prod_{i=1}^3 \int_{-\pi}^\pi \frac{dp_i}{2\pi}\sum_n\left\{-
2\ln\left(2\lambda\right)
+\ln \left(g^2 + h\right)\right\}
\ee
where
\be
g=\lambda+w+n(\mu,\omega_n)n^*(-\mu,\omega_n)(\lambda-w)-
\left[n(\mu,\omega_n)+n^*(-\mu,\omega_n)\right]\sin\omega_n,
\ee
and
\be
h=  \left[n(\mu,\omega_n)-n^*(-\mu,\omega_n)\right]^2
 \sum_{k=1}^3 \sin^2 p_k.
\ee 
The quark number susceptibility is
\be
\chi(N_T,m_w) = 
\frac{1}{N_T}\prod_{i=1}^3 \int_{-\pi}^\pi \frac{dp_i}{2\pi} \sum_n
\frac{4 (q^2-r^2)\left(\sin^2\omega_n - \sum_{k=1}^3
   \sin^2p_k\right) -4s(\omega_n)(\lambda+w)\sin\omega_n}{(\lambda+w)^2},
\ee 
where
\be
q(\omega_n)=\frac{\partial
  n_i(\mu,\omega_n)}{\partial \mu}\Bigg|_{\mu=0};\ \ \ \
r(\omega_n)=\frac{\partial
 n_r(\mu,\omega_n)}{\partial \mu}\Bigg|_{\mu=0};\ \ \ \
s(\omega_n)=\frac{\partial^2
 n_r(\mu,\omega_n)}{\partial^2 \mu}\Bigg|_{\mu=0}.
\ee
The sum over $n$ can be non-zero only if $4(q^2-r^2)$ is an even
function of
$\omega_n$
and $s$ is an odd function of $\omega_n$. 

Two choices one typically makes are
\be
\hat\mu_R(\mu) = \frac{i}{2m_w} \cases { \mu & for n\"aive
  insertion;\cr
\frac{(e^\mu-1)T_4-(e^{-\mu}-1)T_4^\dagger}{2} & for H-K insertion;\cr}
\ee
resulting in
\be
n(\mu,\omega_n) =\frac{1}{2m_w} \cases {i\mu & for n\"aive insertion; \cr \sin\omega_n
-\sin(\omega_n-i\mu) 
& for H-K insertion.\cr}
\ee
The $\frac{1}{m_w}$ factor comes from the tree-level wavefunction
renormalization~\cite{Edwards:1998wx}.
For the n\"aive insertion,
\be
q=\frac{1}{2 m_w}; \ \ \ \ r=0;\ \ \ \ s=0.
\ee
For the H-K insertion,
\be
q=\frac{\cos\omega_n}{2m_w};\ \ \ 
r=0;\ \ \ \
s=-\frac{\sin\omega_n}{2m_w}.
\ee

We can make a modification to the H-K insertion such that $s=0$ but keep
$q$ to be the same. This corresponds to
\be
\hat\mu_R(\mu) = \frac{i}{2m_w} \sinh\frac{\mu}{2} (T_4+T_4^\dagger)\Rightarrow
n(\mu,\omega_n) = \frac{i}{m_w} \cos\omega_n \sinh\frac{\mu}{2}.
\ee
Keeping only the forward derivative will result in
\be
n(\mu,\omega_n) = \frac{i}{m_w} e^{i\omega_n} \sinh\frac{\mu}{2},
\ee
with
\be
q=\frac{\cos\omega_n}{2m_w};\ \ \ 
r=-\frac{\sin\omega_n}{2m_w};\ \ \ 
s=0.
\ee

This leads us to consider the restricted class of operators,
\be
\hat\mu^j_R(\mu) = \frac{i}{m_w} \sinh\frac{\mu}{2} T^j_4\Rightarrow
n^j(\mu,\omega_n) = \frac{i}{m_w} e^{ij\omega_n} \sinh\frac{\mu}{2},
\ee
with
\be
q=\frac{\cos j\omega_n}{2m_w};\ \ \  
r=-\frac{\sin j\omega_n}{2m_w};\ \ \  
s=0;\ \ \ \ q^2-r^2=\frac{\cos 2j \omega_n}{(2m_w)^2}.
\ee
The operator $T_4^j$ is the j-th power of the operator $T_4$ defined 
in (\ref{t4op}). 
This motivates us to study
\be
I_j(N_T,m_w)=
\frac{1}{N_T m_w^2}\prod_{i=1}^3 \int_{-\pi}^\pi \frac{dp_i}{2\pi} \sum_n
\frac{\cos (j\omega_n) \left(\sin^2\omega_n - \sum_{k=1}^3
   \sin^2p_k\right) }
{(\lambda+w)^2},~j=even.
\ee
and we conjecture that 
\be
I_j(N_T,m_w) = I_j(\infty,m_w) + \frac{1}{3N_T^2} +
O\left(N_T^{-4}\right).\label{conj1}
\ee

\begin{figure}
\centerline{\includegraphics[width=\textwidth]{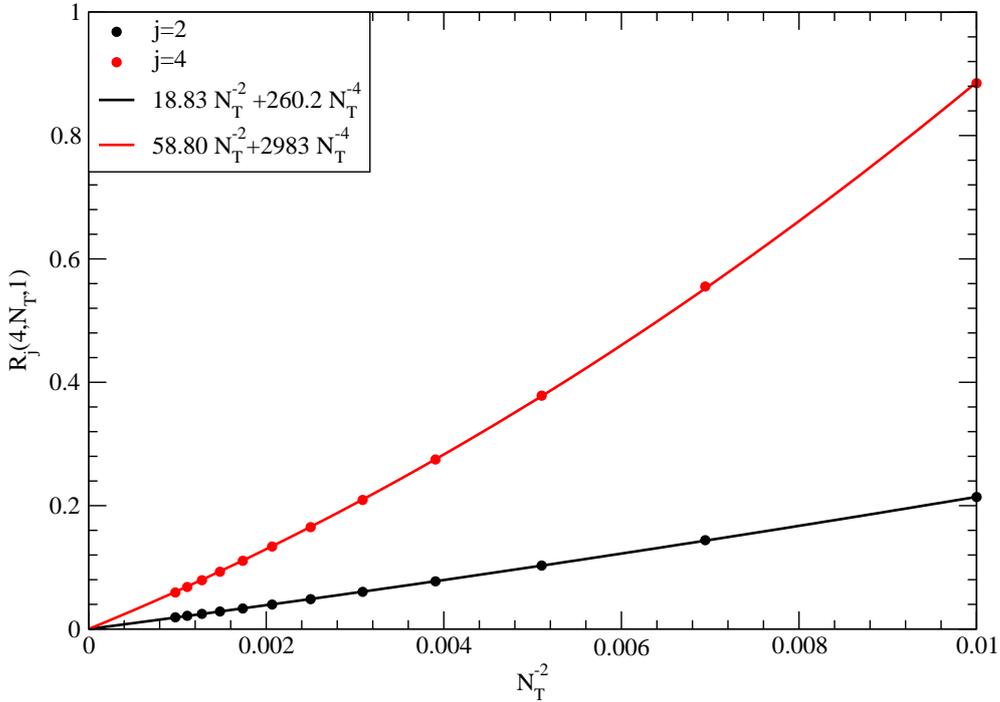}}
\caption{Numerically obtained plot in support of the conjecture.
}
\label{fig1}
\end{figure}

\begin{figure}
\centerline{\includegraphics[width=\textwidth]{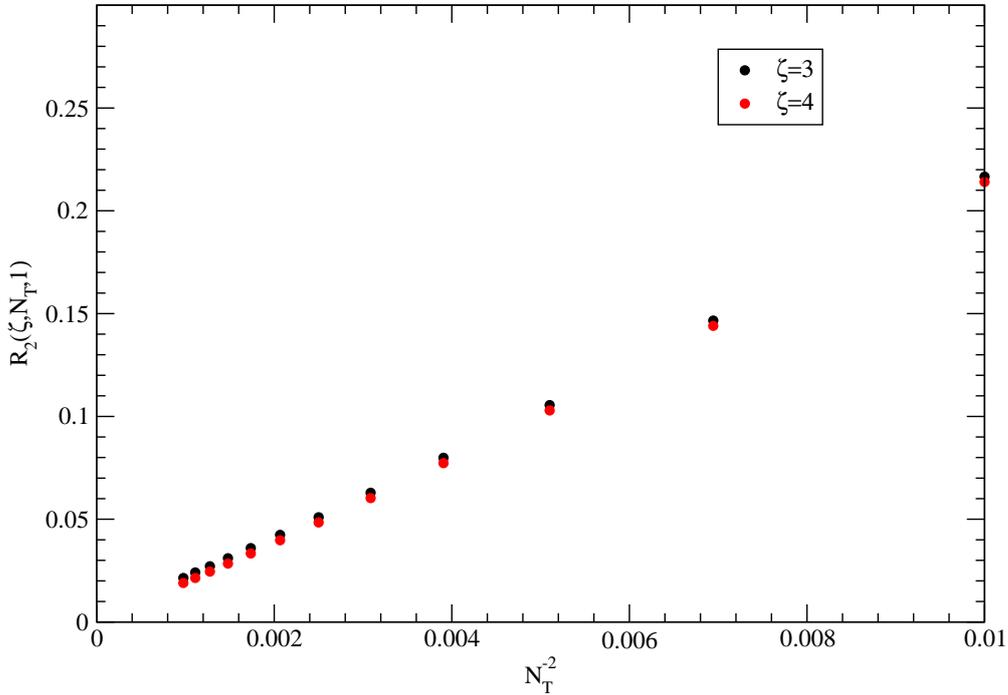}}
\caption{Numerical evidence that $\zeta=4$ is large enough for
the computation at hand.
}
\label{fig2}
\end{figure}

\begin{figure}
\centerline{\includegraphics[width=\textwidth]{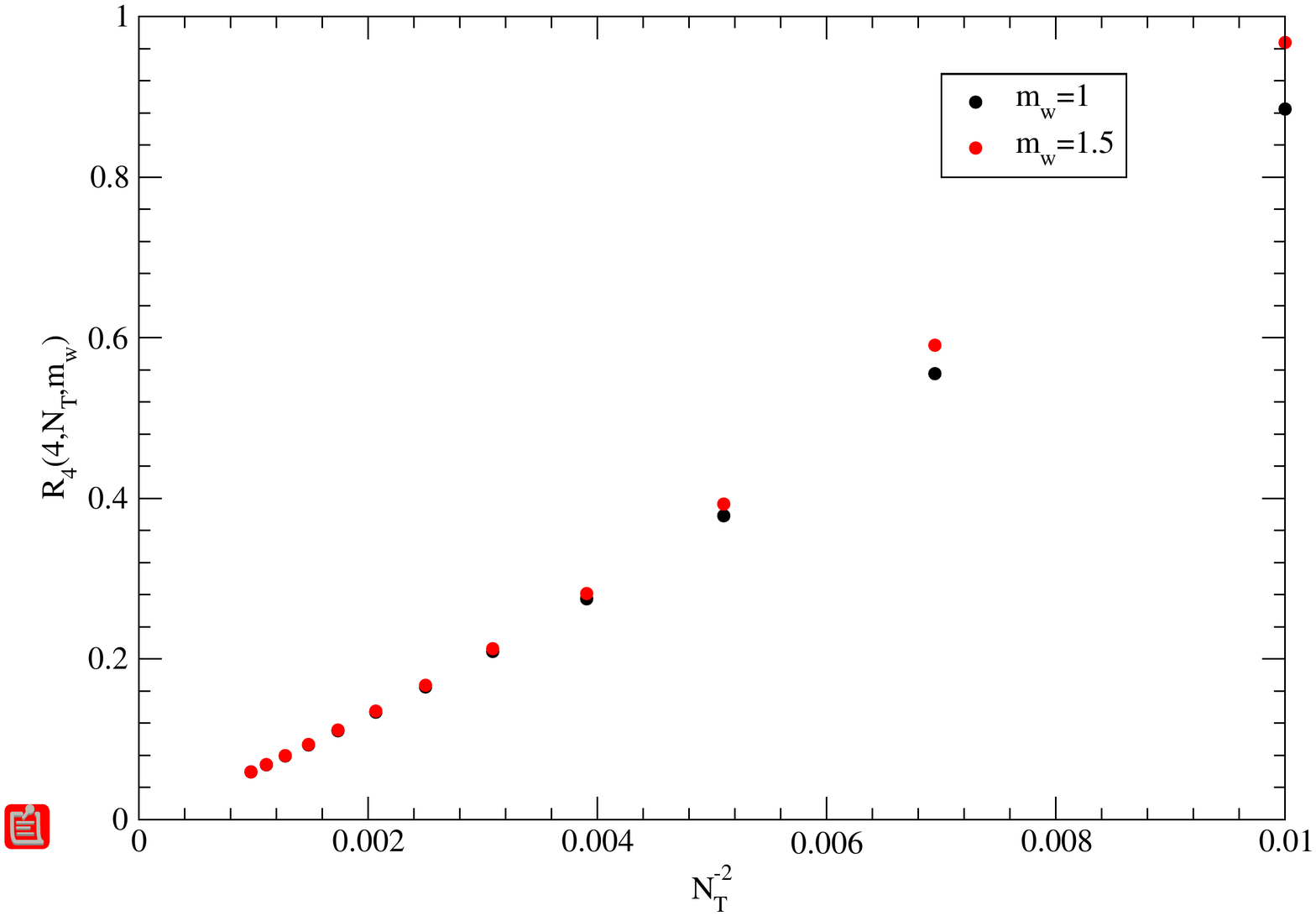}}
\caption{The effect of the Wilson mass parameter on the subleading terms is minimal.
}
\label{fig3}
\end{figure}

We are unable to prove (\ref{conj1}) analytically since we need to
work at a finite but large $N_T$. The resulting contour integral is
complicated due to the presence of the branch cuts associated with
square roots signs in definition of $\lambda$. We resort to a
numerical
check of the conjecture. For this purpose, we convert the three
dimensional
integral over the momenta to a three dimensional sum with $N$ points
in each direction. Let $I_j(N,N_T,m_w)$ denote this sum.
Keeping $N$ fixed, we compute the $I_j(N,\infty,m_w)$ by computing the
quantity for a very large $N_T$. We take this result and extrapolate
to $N\to\infty$ and obtain $I_j(\infty,m_w)$. We then consider
$I_j(\zeta N_T, N_T, m_w) - I_j(\infty,m_w)$ with $\zeta=3,4$ 
in order to extract the $\frac{1}{N_T^2}$ coefficient in (\ref{conj1}).
We numerically evaluated
\be
R_j(\zeta,N_T,m_w) = 3N_T^2 \left( I_j(\zeta N_T, N_T, m_w) -
  I_j(\infty,m_w)\right) -1
\ee
for $j=2,4$ at $\zeta=3,4$ and $m_w=1,1.5$.
It should approach zero with corrections of the order of $\frac{1}{N_T^2}$.
Fig.~\ref{fig1} shows numerical evidence in support of the conjecture:
The fits show that $R_j(4,N_T,1)$ approach zero as $\frac{1}{N_T^2}$
but the corrections get larger with $j$.
Fig.~\ref{fig2} shows that $\zeta=4$ is sufficiently large for the
numerical
computation at hand and Fig.~\ref{fig3} shows that the effect of the
Wilson mass, after taking into account the tree level wavefunction
renormalization factor, is minimal.

\section{Conclusions}
We have discussed the problem of introducing the chemical potential
within the
overlap formalism. The formalism presented here couples the chemical
potential
only to the physical chiral fermions and the overlap definition of the
topological
charge is unaffected by the value of the chemical potential on the
lattice.
We have discussed a large class of operators coupled to the chemical
potential.
Each one of them results in a divergent quark number susceptibility but
all of them have the correct continuum limit after the divergent
contribution is subtracted. We envision using the formalism developed
in this paper in two ways:
\begin{enumerate}
\item Since the main physics aim in the Taylor
expansion method of studying the phase transition from the hadronic
phase
to the quark-gluon plasma phase is to look for growing fluctuations
close
to the phase transition, we could consider, say the quark number
susceptibility, by taking the difference between its value
on a finite temperature lattice and zero temperature lattice for the
same value of lattice gauge couplings and lattice quark
masses~\cite{Gavai:2009vb}.
This will remove the zero temperature divergences and enable a proper 
study of the fluctuations close to the transition temperature.
 The chemical potential was coupled to only those physical fermions
that are confined to the 4D domain wall at the origin of the fifth
dimension in~\cite{Gavai:2009vb}. However the number density term did
not commute 
with the overlap Hamiltonian. 
In this paper the chemical potential is coupled to the conserved number
density operator explicit
in the many body overlap formalism.
\item Consider the linear combination,
\be
\hat\mu_R(\mu) = \sum_{j=1}^\infty c_j \hat\mu_R^j(\mu).
\ee
The associated free fermion quark number susceptibility for this choice of the
number 
density operator will be
\be
\chi(N_T,m_w)= 
\sum_{j_1,j_2=1}^\infty c_{j_1} c_{j_2}
I_{j_1+j_2}(N_T,m_w).
\ee
As per the conjecture in (\ref{conj1}), the quark number
susceptibility will give the correct $\frac{1}{3N_T^2}$ behavior as
long as
\be
\sum_j c_j =1.
\ee
In order to have no divergence, we need to choose the coefficients, $c_j$, such that
\be
\sum_{j_1,j_2=1}^\infty c_{j_1} c_{j_2}
I_{j_1+j_2}(\infty,m_w) =0,
\ee
and this is only one condition and we have infinite coefficients.
Therefore, we can find a large class of number density operators
that have the correct finite behavior in the free fermion limit.
Addition of gauge fields cannot give rise to new divergences as long
as all couplings and masses are properly renormalized. We could
therefore choose $c_1=\alpha$ and $c_2=(1-\alpha)$ with $\alpha$
chosen to cancel the free fermion divergence. This will provide
one good choice for the number density operator in the full interacting
theory. There is clearly one weakness to this approach since one has
to tune $\alpha$ to cancel the divergence. 
\end{enumerate}

Dynamical simulations of overlap fermions are currently being
performed both at zero temperature~\cite{Hashimoto:2008fc} and finite
temperature~\cite{Cossu:2010rc}.
It would be interesting to compute the quark number susceptibility
in such simulations using the operators presented in this paper
and compare them with the complimentary approach
of~\cite{Bloch:2006cd}. The two approaches are quite different in
their construction. We only couple the chemical potential to the physical
fermions. In addition to coupling the chemical potential to the
physical fermions, it is also
coupled to an infinite number of regulator
fields in the Bloch-Wettig scheme. A related issue is the definition
of the topological charge. It does not depend on the chemical
potential
in our scheme contrary to the Bloch-Wettig scheme. The continuum
definition
of the topological charge based on the counting of the zero modes of
the chiral Dirac operator is not affected by the insertion of the
chemical
potential and our scheme maintains this continuum property on the lattice.
Since we expect to
work
with large values of lattice chemical potential particularly close to the
physical
transition that separates hadronic matter from quark-gluon plasma, the
lattice
spacing effects arising from the topological charge depending on the
chemical
potential could be severe. 
In spite of these differences, we expect
lattice simulations to show that the two approaches agree in the
continuum limit.

\appendix
\section{Details of the Hermitian Wilson-Dirac operator}\label{ap1}
\be
H_w = \pmatrix{
B - m_w & C \cr C^\dagger & -B + m_w\cr}
\ee
be the massive Dirac operator with $0 < m_w < 2$. 
In the above equation,
\bea
B &=& \frac{1}{2}\sum_\mu \left( 2 - T_\mu - T_\mu^\dagger\right)\cr
C &=& \frac{1}{2}\sum_\mu \sigma_\mu \left(T_\mu - T_\mu^\dagger\right)\cr
(T_\mu \phi)(x) &=& U_\mu(x) \phi(x+\hat\mu)
\eea
with
\be
\sigma_1=\pmatrix{ 0 & 1 \cr 1 & 0 \cr };\ \ 
\sigma_2=\pmatrix{ 0 & -i \cr i & 0 \cr };\\
\sigma_3=\pmatrix{ 1 & 0 \cr 0 & -1 \cr };\ \ 
\sigma_4=\pmatrix{ i & 0 \cr 0 & i \cr }
\ee
in $d=4$ and
\be
\sigma_1 = 1;\ \ \sigma_2= -i
\ee
in $d=2$. The Dirac matrices in the chiral basis are
\be
\gamma_\mu = \pmatrix{0 & \sigma_\mu \cr \sigma_\mu^\dagger & 0 \cr};\
\ \ \ \gamma_5=\pmatrix{1 & 0 \cr 0 & -1 \cr}.
\ee
We will assume that $H_w$ is a $2n\times 2n$ matrix
and we will assume the gauge field is in the zero topological
sector.\footnote{The overlap formalism is designed to work in all
 topological sectors and this is one of the important features of the
formalism. But, we can restrict ourselves to the zero topological
sector
to simplify the discussion. All results will trivially extend to all
topological sectors.}
Let 
\be
H_w U = U \Lambda,\label{Hweig}
\ee
with
\be
U = \pmatrix{ \alpha & \gamma \cr \beta & \delta\cr};\ \ \ \ 
\Lambda = {\rm diag} (\lambda^+_1,\cdots,\lambda^+_n,
-\lambda^-_1,\cdots, -\lambda^-_n
).\label{umat}
\ee
and $\lambda^\pm_i > 0$ for all $i$.

\section {Details of the derivation of generating functional for
  massive overlap fermions with a chemical potential}\label{ap2}

 Like in (\ref{chemsec}), we can introduce Grassmann variables,
$\zeta_R,\zeta_L,\bar\zeta_R,\bar\zeta_L$, that anticommute
with all fermionic operators and Grassmann variables to rewrite
(\ref{genfunmc1}) in terms of (\ref{genchem}) as
\bea
Z(\bar\xi,\xi;\hat\mu_R,\hat\mu_L;m) &=&
\int d\zeta_L d \bar\zeta_R d\bar\zeta_L d\zeta_R e^{\bar\zeta_R \zeta_L -\bar\zeta_L\zeta_R}\cr
&&\left[ {}_R\langle-|\otimes {}_L\langle +| \right]
e^{- \sqrt{m}\bar\zeta_R d_R
   -\sqrt{m}\zeta_R u_R^\dagger
   -\sqrt{m}\bar\zeta_L u_L -\sqrt{m}\zeta_L d^\dagger_L}\cr
&&e^{\bar\xi_R d_R+ \xi_R u^\dagger_R+ u^\dagger_R
\hat\mu_R d_R +\xi_L d^\dagger_L  +\bar\xi_L u_L - d^\dagger_L
\hat\mu_L u_L}\left[ |-\rangle_L \otimes |+\rangle_R \right].
\eea
Using (\ref{genchem1}) the above equation can be written as 
\bea
Z(\bar\xi,\xi;\hat\mu_R,\hat\mu_L;m)
&=&
\det \alpha \det \left( 1+ \hat\mu_R \beta\alpha^{-1}\right)
\det \alpha^\dagger \det \left( 1 -\hat\mu_L
\left(\beta\alpha^{-1}\right)^{\dagger}\right) \cr
&& \int d\zeta_L d\bar\zeta_R d\bar\zeta_L d\zeta_R e^{\bar\zeta_R \zeta_L -\bar\zeta_L \zeta_R}\cr
&&
e^{\left(\bar\xi_R -\sqrt{m}\bar\zeta_R\right)
\frac{1}{\alpha\beta^{-1}+\hat\mu_R}
\left(\xi_R-\sqrt{m}\zeta_R\right)}\cr
&&e^{\left(\bar\xi_L-\sqrt{m}\bar\zeta_L\right)
\left[ \frac{1}{\left(\alpha\beta^{-1}\right)^\dagger-\hat\mu_L}\right]
\left(\xi_L-\sqrt{m}\zeta_L\right)}.
\eea
We can write the exponent in the integrand as
\be
S= \bar\zeta F \zeta - \sqrt{m} \left( \bar\zeta G \xi +\bar\xi G
 \zeta\right) + \bar\xi G \xi
\ee
where
\be
\bar\zeta=\pmatrix{\bar\zeta_R & \bar\zeta_L\cr};\ \ \ 
\zeta=\pmatrix{\zeta_R \cr \zeta_L\cr};
\ee
and
\be
F=\pmatrix{ mG_1 & 1 \cr -1 & mG_2^\dagger \cr};\ \ \ 
G=\pmatrix{G_1 & 0 \cr 0 & G_2^\dagger\cr},
\ee
and
\be
G_1 =\frac{1}{\alpha\beta^{-1} + \hat\mu_R};\ \  
G_2^{\dagger} =\frac{1}{(\alpha\beta^{-1})^{\dagger} - \hat\mu_L}.
\ee 
Integration of $\bar\zeta$ and $\zeta$ yields 
\be
\det F e^{\bar\xi \left ( G - mGF^{-1}G\right) \xi}.
\ee
The second line in the last equality of (\ref{genfunmc}) is $\det F$.
One can show that
\be
F^{-1} = \pmatrix{ mG_2^\dagger \frac{1}{1+ m^2 G_1 G_2^\dagger} &
   -\frac{1}{1+m^2 G_2^\dagger G_1} \cr
\frac{1}{1+m^2 G_1 G_2^\dagger} & mG_1 \frac{1}{1+m^2 G_2^\dagger
 G_1}\cr},
\ee
and it follows that
\be
G-mG F^{-1} G = \pmatrix {
\frac{1}{1+m^2G_1G_2^\dagger}G_1 &
G_1 \frac{m}{1+m^2 G_2^\dagger G_1} G_2^\dagger \cr
-G_2^\dagger\frac{m}{1+m^2G_1G_2^\dagger}G_1 &
\frac{1}{1+m^2 G_2^\dagger G_1} G_2^\dagger \cr
}
\ee
from which the last two lines of (\ref{genfunmc}) follows.

\acknowledgments

R.N. would like to thank the Theory Group at TIFR for two invitations
in the past two years that triggered the current collaboration. The
authors
would like to acknowledge several useful discussions with Rajiv Gavai.
R.N. acknowledges partial support by the NSF under grant number
PHY-0854744.


\begin{thebibliography}{99}
\bibitem{Endrodi:2011gv}
  G.~Endrodi, Z.~Fodor, S.~D.~Katz, K.~K.~Szabo,
  JHEP {\bf 1104}, 001 (2011).
[arXiv:1102.1356 [hep-lat]].
\bibitem{Gupta:2009mu}
  S.~Gupta,
  PoS {\bf CPOD2009}, 025 (2009).
  [arXiv:0909.4630 [nucl-ex]].
\bibitem{Gavai:2008zr}
  R.~V.~Gavai, S.~Gupta,
  Phys.\ Rev.\  {\bf D78}, 114503 (2008).
  [arXiv:0806.2233 [hep-lat]].
\bibitem{Ejiri:2005wq}
 S.~Ejiri, F.~Karsch, K.~Redlich,
 Phys.\ Lett.\  {\bf B633}, 275-282 (2006).
 [hep-ph/0509051].
\bibitem{Allton:2002zi}
  C.~R.~Allton, S.~Ejiri, S.~J.~Hands, O.~Kaczmarek, F.~Karsch, E.~Laermann, C.~Schmidt, L.~Scorzato,
  Phys.\ Rev.\  {\bf D66}, 074507 (2002).
  [hep-lat/0204010].
\bibitem{Gavai:2003mf}
  R.~V.~Gavai, S.~Gupta,
  Phys.\ Rev.\  {\bf D68}, 034506 (2003).
  [hep-lat/0303013].
\bibitem{DeTar:2010xm}
  C.~DeTar, L.~Levkova, S.~Gottlieb, U.~M.~Heller, J.~E.~Hetrick, R.~Sugar, D.~Toussaint,
  Phys.\ Rev.\  {\bf D81}, 114504 (2010).
  [arXiv:1003.5682 [hep-lat]].
\bibitem{Asakawa:2000wh}
  M.~Asakawa, U.~W.~Heinz, B.~Muller,
  Phys.\ Rev.\ Lett.\  {\bf 85}, 2072-2075 (2000).
  [hep-ph/0003169].
\bibitem{Gavai:2005rf}
  R.~V.~Gavai, S.~Gupta,
  Eur.\ Phys.\ J.\  {\bf C43}, 31-34 (2005).
  [hep-ph/0502198].
\bibitem{Blaizot:2002xz}
  J.~P.~Blaizot, E.~Iancu, A.~Rebhan,
  Eur.\ Phys.\ J.\  {\bf C27}, 433-438 (2003).
  [hep-ph/0206280].
\bibitem{Hasenfratz:1983ba}
  P.~Hasenfratz, F.~Karsch,
  Phys.\ Lett.\  {\bf B125}, 308 (1983).
\bibitem{Neuberger:1997fp}
 H.~Neuberger,
 Phys.\ Lett.\  B {\bf 417}, 141 (1998)
 [arXiv:hep-lat/9707022].
\bibitem{Bloch:2006cd}
  J.~C.~R.~Bloch and T.~Wettig,
  Phys.\ Rev.\ Lett.\  {\bf 97}, 012003 (2006)
  [arXiv:hep-lat/0604020].
\bibitem{Bloch:2007xi}
  J.~C.~R.~Bloch and T.~Wettig,
  Phys.\ Rev.\  D {\bf 76}, 114511 (2007)
  [arXiv:0709.4630 [hep-lat]].
\bibitem{Neuberger:1997bg}
  H.~Neuberger,
  Phys.\ Rev.\  {\bf D57}, 5417-5433 (1998).
  [hep-lat/9710089].
\bibitem{Edwards:1998wx}
  R.~G.~Edwards, U.~M.~Heller, R.~Narayanan,
  Phys.\ Rev.\  {\bf D59}, 094510 (1999).
  [hep-lat/9811030].
\bibitem{Gattringer:2007uu}
  C.~Gattringer, L.~Liptak,
  Phys.\ Rev.\  {\bf D76}, 054502 (2007).
  [arXiv:0704.0092 [hep-lat]].
\bibitem{Banerjee:2008ii}
  D.~Banerjee, R.~V.~Gavai and S.~Sharma,
  Phys.\ Rev.\  D {\bf 78}, 014506 (2008)
  [arXiv:0803.3925 [hep-lat]].
\bibitem{Gavai:2009vb}
  R.~V.~Gavai, S.~Sharma,
  Phys.\ Rev.\  {\bf D81}, 034501 (2010).
  [arXiv:0906.5188 [hep-lat]].
\bibitem{Narayanan:1994gw}
  R.~Narayanan and H.~Neuberger,
  Nucl.\ Phys.\  B {\bf 443}, 305 (1995)
  [arXiv:hep-th/9411108].
\bibitem{Hashimoto:2008fc}
  S.~Hashimoto,
  PoS {\bf LATTICE2008}, 011 (2008).
  [arXiv:0811.1257 [hep-lat]].
\bibitem{Cossu:2010rc}
  G.~Cossu {\it et al.} [ JLQCD Collaboration ],
  PoS {\bf LATTICE2010}, 174 (2010).
  [arXiv:1011.0257 [hep-lat]].
\end{thebibliography}
\end{document}